%% file: paper.tex
\documentclass[letterpaper, 10 pt, conference]{ieeeconf}  

\IEEEoverridecommandlockouts                              

\usepackage{graphicx,xcolor}      
\usepackage{amsmath, amssymb}
\usepackage{cuted}
\usepackage{tikz}

\tikzset{>=stealth}
\usetikzlibrary{arrows,shapes,calc}
\usetikzlibrary{positioning}
\usetikzlibrary{external}
\usetikzlibrary{decorations.pathreplacing,decorations.markings}

\tikzset{%
	mid arrow/.style={postaction={decorate,decoration={				markings,mark=at position .5 with {\arrow[#1]{latex}}}}},
	mid arrow reversed/.style={postaction={decorate,decoration={				markings,mark=at position .5 with {\arrowreversed[#1]{latex}}}}},
	block/.style    = {draw, thick, rectangle, minimum height = 3em,
		minimum width = 2.5em, node distance = 6em, colorODE},
	dots/.style    = {draw = none},
	PDE/.style    = {draw, thick, rectangle, minimum height = 3em,
		minimum width = 8em, node distance = 6em, colorPDE},
	sum/.style      = {draw, circle, node distance = 2cm},
	input/.style    = {coordinate},
	output/.style   = {coordinate},
	gain/.style     = {draw, isosceles triangle, anchor = west,shape border rotate =-90, inner sep = 1pt, minimum size = 3em},
	box/.style    = {draw, thick, rectangle, minimum height = 3em,
		minimum width = 6.9em, text width = 6em, align = center, node distance = 6em, color3, fill=color3!5}
}

\definecolor{colorODE}{RGB}{200,200,200}
\definecolor{colorPDE}{RGB}{253,127,0}
\definecolor{colorInput}{RGB}{0,127,253}

\newcommand{\dd}{\mathrm{d}}
\newcommand{\Rset}{\mathbb{R}}

\newcommand{\Nset}{\mathbb{N}}
\newcommand{\tint}{\textstyle\int}
\DeclareMathOperator{\rank}{rk}

\newtheorem{lemma}{Lemma}
\newtheorem{definition}{Definition}
\newtheorem{remark}{Remark}
\newtheorem{assumption}{Assumption}

\title{\LARGE \bf On the strict-feedback form of hyperbolic distributed-parameter systems}

\author{Nicole Gehring
\thanks{
	Nicole Gehring is with the Chair of Systems Theory and Control Engineering, Otto von Guericke University Magdeburg, 39106 Magdeburg, Germany {\tt\small nicole.gehring@ovgu.de}}%
}

\begin{document}

\maketitle
\thispagestyle{empty}
\pagestyle{empty}

\begin{abstract}
The paper is concerned with the strict-feedback form of hyperbolic distributed-parameter systems.
Such a system structure is well known to be the basis for the recursive backstepping control design for nonlinear ODEs and is also reflected in the Volterra integral transformation used in the backstepping-based stabilization of parabolic PDEs.
Although such integral transformations also proved very helpful in deriving state feedback controllers for hyperbolic PDEs, they are not necessarily related to a strict-feedback form.
Therefore, the paper looks at structural properties of hyperbolic systems in the context of controllability.
By combining and extending existing backstepping results, exactly controllable heterodirectional hyperbolic PDEs as well as PDE-ODE systems are mapped into strict-feedback form.
While stabilization is not the objective in this paper, the obtained system structure is the basis for a recursive backstepping design and provides new insights into coupling structures of distributed-parameter systems that allow for a simple control design.
In that sense, the paper aims to take backstepping for PDEs back to its ODE origin.
\end{abstract}

\section{Introduction}

The term strict-feedback form typically refers to a special state representation of a nonlinear system.
In the single-input case, \cite{Sastry1999} introduces the strict-feedback form
\begin{subequations}
	\label{eq:ode-sff}
	\begin{align}
		\dot x_1 &= f_1(x_1) + x_2 \\
		\dot x_2 &= f_2(x_1,x_2) + x_3 \\
		&\,\,\,\vdots \nonumber \\
		\dot x_n &= f_n(x_1,\dots,x_n) + u
	\end{align}
\end{subequations}
with state $x(t) = [x_1(t),\dots,x_n(t)]^T\in\Rset^n$ and input $u(t)\in\Rset$.
Though there exist different definitions, they all share that the $i$-th differential equation, $i=1,\dots,n$, only depends on state components $x_1,\dots,x_i$ as well as affinely on $x_{i+1}$, with $x_{n+1}=u$.
A system of the form \eqref{eq:ode-sff} is well-known to be differentially flat, with a flat output $y=x_1$ (e.g., \cite{Fliess1995}).
This property coincides with controllability in the linear case.
The triangular form of \eqref{eq:ode-sff} is exploited in the recursive backstepping control design (e.g., \cite{Krstic1995}), where in the $i$-th step, $x_{i+1}$ takes the role of a virtual input.


The backstepping design of controllers for distributed-parameter systems makes use of a Volterra integral transformation to map a given dynamics into a target system, where the choice of a stabilizing state feedback is obvious.
The popular approach originated from \cite{Balogh2002ejc}, where the backstepping design for nonlinear ordinary differential equations (ODEs) is transferred to infinite-dimensional systems by applying a spatial discretization to a reaction-diffusion equation with boundary input in order to obtain a system in strict-feedback form.
Tracing the result of the recursive control design back to a Volterra integral transformation for the original spatially distributed state, this introduced backstepping for distributed-parameter systems.
By a spatial discretization of the integral transformation, the triangular form inherent to backstepping for nonlinear ODEs is apparent.
In that sense, a boundary-actuated parabolic partial differential equation (PDE) can be considered to be in strict-feedback form (see also \cite{Krstic2008book} for a link between strict-feedback systems and backstepping).

Backstepping or Volterra integral transformations are also used to derive stabilizing controllers for heterodirectional hyperbolic PDEs.
In \cite{Vazquez2011CDC} and \cite{Coron2013SIAM}, $2\times2$ systems
\begin{subequations}
	\label{eq:hPDE-SISO}
	\begin{align}
		\partial_t x^-(z,t) &= \lambda^-(z)\partial_z x^-(z,t) + a^{-+}(z)x^+(z,t) \\
		\partial_t x^+(z,t) &= -\lambda^+(z)\partial_z x^+(z,t) + a^{+-}(z)x^-(z,t) \\
		\label{eq:hPDE-SISO-bc0}
		x^+(0,t) &= q x^-(0,t) \\
		x^-(1,t) &= r x^+(1,t) + u(t)
	\end{align}
\end{subequations}
with state $x(z,t)=[x^-(z,t),x^+(z,t)]^T\in\Rset^2$ and input $u(t)\in\Rset$ are considered, where the components $x^-(z,t)$ and $x^+(z,t)$ propagate in opposite directions.
The kernel $K(z,\zeta)\in\Rset^{2\times2}$ of a Volterra integral transformation
\begin{equation}
	\label{eq:vit}
	\bar x(z,t) = x(z,t) - \tint_0^z K(z,\zeta)x(\zeta,t)\,\dd\zeta
\end{equation}
is determined depending on whether $q\neq0$ or $q=0$ in \eqref{eq:hPDE-SISO-bc0}.
This distinction corresponds to whether or not the hyperbolic system \eqref{eq:hPDE-SISO} is exactly controllable or not (see, e.g., \cite{Russell1978siam,Coron2019siam}).
If $q=0$, the system is still null controllable, meaning that stabilization of $x^-(z,t)$ is sufficient to ensure convergence of $x^+(z,t)$ to zero.
Different backstepping controllers have been designed that guarantee stabilization of general classes of hyperbolic systems (see \cite{Hu2016tac,Auriol2016AUT,Coron2017AUT,Hu2019siam,Coron2019siam}).

Here, the main focus is not on the design of stabilizing feedback controllers.
Instead, the contribution aims to bring backstepping for heterodirectional hyperbolic systems back to its ODE origins.
Looking at a spatial discretization of \eqref{eq:vit}, in view of the two components of $x(z,t)$, it is apparent that such a transformation does not possess the triangular structure that is typical in backstepping for both nonlinear ODEs and parabolic PDEs.
Thus, backstepping for hyperbolic systems currently does not build on a strict-feedback form of the PDEs themselves.
Still, there are several results that either explicitly or implicitly make use of just such a form to design controllers for interconnected infinite-dimensional systems involving both PDEs and ODEs (e.g., \cite{Gehring2021MTNS,Auriol2021ecc,Xu2023aut}).
Moreover, both for nonlinear ODEs and parabolic PDEs, there is a direct relation between the flatness of these systems and the existence of a strict-feedback form.
In \cite{Gehring2023at}, this link is also shown to exist for linear PDE-ODE single-input systems (see also \cite{Woittennek2012mathmod}).
Importantly, the flatness of such a system requires exact controllability of \eqref{eq:hPDE-SISO}.

This paper introduces a strict-feedback form for heterodirectional hyperbolic systems with multiple inputs as well as hyperbolic PDE-ODE systems.
For that, it has to be assumed that the PDE (sub)system is exactly controllable.
Starting from a standard representation of the dynamics, the system is systematically transformed into strict-feedback form.
In that, new kernel equations arise based on those in \cite{Hu2019siam}, for which available degrees of freedom and the property of exact controllability are explicitly used.
To the best of the author's knowledge, so far, the controllability property has only been taken into account in determining the kernel of Volterra integral transformations in the single-input case.
In the end, the strict-feedback form is shown to be the basis for extending the recursive design of backstepping controllers to larger classes of systems.
The form also gives new insight into the coupling structure of distributed-parameter systems.

\textit{Structure:} The heterodirectional hyperbolic system and its properties are discussed in Section~\ref{sec:problem}.
In Section~\ref{sec:transformation}, the system is mapped into strict-feedback form by two successive transformations.
Similarly, in Section~\ref{sec:po}, a PDE-ODE system is transformed into strict-feedback form.

\textit{Notation:} Denote by $I_k$ the identity matrix in $\Rset^{k\times k}$ for some $k\in\Nset$.
The notation $[M]_\ast$ means that only those elements $M_{ij}$ of a matrix $M$ in the $i$-th row and $j$-th column are considered that satisfy the condition $\ast$.
The use of $\pm$ as a sub- or superscript means that a statement applies to both cases, $+$ and $-$ (e.g., $x^\pm\in\Rset^{n_\pm}$ for $x^+\in\Rset^{n_+}$, $x^-\in\Rset^{n_-}$).

\section{Problem statement}
\label{sec:problem}

Consider the heterodirectional hyperbolic system
\begin{subequations}
	\label{eq:sys}
	\begin{align}
		\label{eq:sys_PDE}
		\begin{bmatrix} \partial_t x^-(z,t) \\ \partial_t x^+(z,t) \end{bmatrix} &= \begin{bmatrix} \Lambda^-(z) & 0 \\ 0 & -\Lambda^+(z) \end{bmatrix} \begin{bmatrix} \partial_z x^-(z,t) \\ \partial_z x^+(z,t) \end{bmatrix} \\
		&\hspace{1cm} + \begin{bmatrix} A^{--}(z) & A^{-+}(z) \\ A^{+-}(z) & A^{++}(z) \end{bmatrix} \begin{bmatrix} x^-(z,t) \\ x^+(z,t) \end{bmatrix} \nonumber \\
		\label{eq:sys_RB0}
		x^+(0,t) &= Q x^-(0,t) \\
		\label{eq:sys_RB1}
		x^-(1,t) &= R x^+(1,t) + u(t),
	\end{align}
\end{subequations}
where the components in $x^-(z,t)\in\Rset^{n_-}$ of the state $x(z,t)=\text{col}(x^-(z,t),x^+(z,t))\in\Rset^n$, $n=n_-+n_+$, propagate from $z=1$ to $z=0$ and those in $x^+(z,t)\in\Rset^{n_+}$ describe a transport in the opposite direction.
The respecetive velocities $\lambda_i^\pm\in C^1([0,1])$, $i=1,\dots,n_\pm$, in the diagonal matrices $\Lambda^\pm(z)=\mathrm{diag}(\lambda_1^\pm(z),\dots,\lambda_{n_\pm}^\pm(z))$ satisfy
\begin{equation}
	\lambda_1^\pm(z)>\cdots>\lambda_{n_\pm}^\pm(z), \qquad z\in[0,1].
\end{equation}
Without loss of generality, it is assumed that the diagonal elements of $A^{--}(z)$ and $A^{++}(z)$ are zero (cf.\ \cite{Hu2016tac}), with all elements of the four in-domain coupling matrices $A^{\pm\pm}$ piecewise continuous.
In addition, both PDE subsystems, i.e., the dynamics for $x^-(z,t)$ and $x^+(z,t)$, respectively, are interconnected at the boundaries via $Q\in\Rset^{n_+\times n_-}$ and $R\in\Rset^{n_-\times n_+}$.
The control input $u(t)\in\Rset^{n_-}$ fully actuates the boundary at $z=1$.

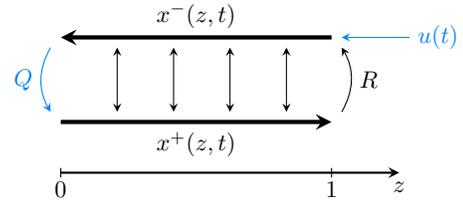
\begin{figure}
	\centering
	\input{figs/fig-system.tex}
	\caption{Coupling structure of the heterodirectional hyperbolic system \eqref{eq:sys}. The arrows in blue (\textcolor{colorInput}{\rule[.5ex]{3ex}{.2ex}}) highlight actions related to controllability.}
	\label{fig:sys}
\end{figure}

The system \eqref{eq:sys} is not in strict-feedback form because of the bidirectional in-domain coupling of both PDE subsystems (see also Fig.~\ref{fig:sys}).
It would be a strict-feedback system if $A^{+-}(z)=0$ .
In order to map \eqref{eq:sys} into strict-feedback form, the following assumption is imposed.

\begin{assumption}
	\label{ass:Q}
	The system \eqref{eq:sys} is exactly controllable, i.e, $\rank Q=n_+$.
\end{assumption}

See \cite{Russell1978siam} and \cite{Coron2019siam} for details on exact controllability.
The assumption implies $n_+\le n_-$.
This is necessary for the existence of a strict-feedback form for \eqref{eq:sys}, as $x^-(0,t)$ takes the role of an input w.r.t.\ the dynamics of $x^+(z,t)$.

\begin{remark}
	\label{rem:stabilizability}
	If $\rank Q<n_+$, \eqref{eq:sys} is (only) null controllable (cf.\ \cite{Coron2019siam}).
	This can be interpreted as part of \eqref{eq:sys} being uncontrollable via $u(t)$ but stable on its own.
	One could also define a strict-feedback form for such systems.
	In fact, for nonlinear ODEs, flatness (or controllability in an appropriate sense) is not implied by definitions of strict-feedback forms that allow $x_1(t)\not\in\Rset$ in \eqref{eq:ode-sff} (e.g., \cite{Krstic1995}).
\end{remark}

The rank condition $\rank Q=n_+$ in Assumption~\eqref{ass:Q} ensures the existence of a right inverse matrix $Q^\text{R}=Q^T(QQ^T)^{-1}\in\Rset^{n_-\times n_+}$ that satisfies $QQ^\text{R}=I_{n_+}$. Moreover, a full-rank right annihilator $Q^\perp\in\Rset^{n_-\times\Delta n}$, $\Delta n=n_--n_+\ge0$, with $QQ^\perp=0_{n_+\times\Delta n}$ exists that results in an invertible matrix
\begin{equation}
	\label{eq:def-T}
	T = \begin{bmatrix} Q^\text{R} & Q^\perp \end{bmatrix}.
\end{equation}
This allows to rewrite the boundary condition \eqref{eq:sys_RB0} as
\begin{equation}
	x^+(0,t) = Q \begin{bmatrix} Q^\text{R} & Q^\perp \end{bmatrix} T^{-1} x^-(0,t) = \begin{bmatrix} I_{n_+} & 0 \end{bmatrix} \tilde x^-(0,t).
\end{equation}
Partitioning
\begin{equation}
	\label{eq:x0tilde}
	\tilde x^-(0,t) = \begin{bmatrix} \tilde x_1^-(0,t) \\ \tilde x_2^-(0,t) \end{bmatrix} = T^{-1} x^-(0,t)
\end{equation}
into parts $\tilde x_1^-(0,t)\in\Rset^{n_+}$ and $\tilde x_2^-(0,t)\in\Rset^{\Delta n}$ then gives $x^+(0,t)=\tilde x_1^-(0,t)$.

In what follows, \eqref{eq:sys} is mapped into strict-feedback form by using Assumption~\ref{ass:Q}.
Although the discussion is done in general, the special case of $\Delta n=0$, i.e., when $Q$ is invertible and $n_+=n_-$, will be of special interest.

\section{Transformation into strict-feedback form}
\label{sec:transformation}

Two transformations\footnote{Note that the transformations in Sections~\ref{sec:vit} and \ref{sec:fit} do more than map the hyperbolic system \eqref{eq:sys} into strict-feedback form. While a different, more minimalist approach is possible, the one chosen here is the most direct.} are used to map the heterodirectional hyperbolic system \eqref{eq:sys} into strict-feedback form.
First, a Volterra integral transformation removes the entire in-domain coupling 
of the PDEs in \eqref{eq:sys_PDE}. 
In contrast to \cite{Hu2019siam}, additional degrees of freedom in the choice of boundary values for the integral kernel are utilized.
In the second and final step, a Fredholm-type integral transformation essentially swaps the order of the PDE subsystems.
Based on the resulting strict-feedback form for \eqref{eq:sys}, more general interconnected hyperbolic systems are discussed in Section~\ref{sec:sff}, for which a control design is particularly simple due to their special coupling structure.

\subsection{Volterra integral transformation}
\label{sec:vit}

Consider the Volterra integral transformation
\begin{equation}
	\label{eq:trafo1}
	\bar x(z,t) = x(z,t) - \tint_0^z K(z,\zeta) x(\zeta,t) \,\dd\zeta,
\end{equation}
where the kernel
\begin{equation}
	K(z,\zeta) = \begin{bmatrix} K^{--}(z,\zeta) & K^{-+}(z,\zeta) \\ K^{+-}(z,\zeta) & K^{++}(z,\zeta) \end{bmatrix} \in \Rset^{n\times n}
\end{equation}
on $\mathcal T=\{(z,\zeta)\in[0,1]^2|\zeta\le z\}$ is partitioned into blocks corresponding to the dimensions of $x^-(z,t)$ and $x^+(z,t)$.
In \cite{Hu2019siam}, the kernel is chosen such the transformed PDE takes the form
\begin{subequations}
	\label{eq:sysHu}
	\begin{align}
		\label{eq:sysHu-m}
		\partial_t \bar x^-(z,t) &= \phantom{-}\Lambda^-(z) \partial_z \bar x^-(z,t) + A_0^-(z) \bar x^-(0,t) \\
		\label{eq:sysHu-p}
		\partial_t \bar x^+(z,t) &= -\Lambda^+(z) \partial_z \bar x^+(z,t) + C_0^+(z) \bar x^-(0,t),
	\end{align}
\end{subequations}
with a strictly lower triangular matrix $A_0^-(z)\in\Rset^{n_-\times n_-}$ and a matrix $C_0^+(z)\in\Rset^{n_+\times n_-}$ without a special form that are both defined in terms of $K(z,\zeta)$.
Instead, here, the transformation will enforce symmetry between both PDE subsystems by taking into account Assumption~\ref{ass:Q}.
In view of \eqref{eq:x0tilde} and $\bar x(0,t)=x(0,t)$, it is possible to write
\begin{equation}
	C_0^+(z) \bar x^-(0,t) = C_0^+(z) Q^\text{R} \bar x^+(0,t) + C_0^+(z)Q^\perp \tilde x_2^-(0,t).
\end{equation}
This gives rise to
\begin{multline}
	\label{eq:temp}
	\partial_t \bar x^+(z,t) = -\Lambda^+(z) \partial_z \bar x^+(z,t) + A_0^+(z) \bar x^+(0,t) \\
	+ B_0^+(z)\tilde x_2^-(0,t),
\end{multline}
which replaces \eqref{eq:sysHu-p} in the transformed system.
Therein, the boundary value $\bar x^+(0,t)$ (contrary to $\bar x^-(0,t)$ in \eqref{eq:sysHu-p}) is multiplied with a strictly lower triangular matrix $A_0^+(z)\in\Rset^{n_+\times n_+}$.
As $\tilde x_2^-(0,t)\in\Rset^{\Delta n}$, the matrix $B_0^+(z)\in\Rset^{n_+\times\Delta n}$ without a special form only exists in the asymmetric case where $\Delta n>0$.

In order for \eqref{eq:trafo1} to map \eqref{eq:sys_PDE} into \eqref{eq:sysHu-m} and \eqref{eq:temp}, $K^{--}(z,\zeta)$ and $K^{-+}(z,\zeta)$ have to satisfy the kernel equations
\begin{subequations}
	\label{eq:trafo1_kerneleq1}
	\begin{align}
		& \Lambda^-(z) \partial_z K^{--}(z,\zeta) + \dd_\zeta(K^{--}(z,\zeta)\Lambda^-(\zeta)) \\
		&\hspace{1cm} = K^{--}(z,\zeta)A^{--}(\zeta) + K^{-+}(z,\zeta)A^{+-}(\zeta) \nonumber\\
		& \Lambda^-(z) \partial_z K^{-+}(z,\zeta) - \dd_\zeta(K^{-+}(z,\zeta)\Lambda^+(\zeta)) \\
		&\hspace{1cm} = K^{--}(z,\zeta)A^{-+}(\zeta) + K^{-+}(z,\zeta)A^{++}(\zeta) \nonumber\\
		& K^{--}(z,z)\Lambda^-(z) - \Lambda^-(z)K^{--}(z,z) = A^{--}(z) \\
		& - K^{-+}(z,z)\Lambda^+(z) - \Lambda^-(z)K^{-+}(z,z) = A^{-+}(z) \\
		\label{eq:trafo1_kerneleq1_bc0}
		& \left[K^{--}(z,0)\Lambda^-(0) - K^{-+}(z,0)\Lambda^+(0)Q\right]_{i\le j} = 0,
	\end{align}
\end{subequations}
with
\begin{subequations}
	\label{eq:trafo1_kerneleq2}
	\begin{align}
		& -\Lambda^+(z)\partial_z K^{++}(z,\zeta) - \dd_\zeta(K^{++}(z,\zeta)\Lambda^+(\zeta)) \\
		&\hspace{1cm} = K^{++}(z,\zeta) A^{++}(\zeta) + K^{+-}(z,\zeta) A^{-+}(\zeta) \nonumber\\
		& -\Lambda^+(z)\partial_z K^{+-}(z,\zeta) + \dd_\zeta(K^{+-}(z,\zeta)\Lambda^-(\zeta)) \\
		&\hspace{1cm} = K^{++}(z,\zeta) A^{+-}(\zeta) + K^{+-}(z,\zeta) A^{--}(\zeta) \nonumber\\
		& - K^{++}(z,z) \Lambda^+(z) + \Lambda^+(z) K^{++}(z,z) = A^{++}(z) \\
		& K^{+-}(z,z)\Lambda^-(z) + \Lambda^+(z) K^{+-}(z,z) = A^{+-}(z) \\
		\label{eq:trafo1_kerneleq2_bc0}
		& \!\left[- K^{++}(z,0)\Lambda^+(0) + K^{+-}(z,0)\Lambda^-(0) Q^\text{R}\right]_{i\le j} = 0\!
	\end{align}
\end{subequations}
for $K^{+-}(z,\zeta)$ and $K^{++}(z,\zeta)$.
The derivation of these equations makes explicit use of Assumption~\ref{ass:Q}.
First, as in \cite{Hu2019siam}, \eqref{eq:trafo1_kerneleq1_bc0} follows from using \eqref{eq:sys_RB0} in
\begin{multline}
	\label{eq:bilanz1}
	K^{--}(z,0)\Lambda^-(0)x^-(0,t) - K^{-+}(z,0)\Lambda^+(0)\underbrace{x^+(0,t)}_{Qx^-(0,t)} \\
	= A_0^-(z) x^-(0,t),
\end{multline}
because in order for this equation to hold for arbitrary values of $x^-(0,t)$, the coefficient of $x^-(0,t)$ has to be zero.
Symmetrically, the balancing is done w.r.t.\ $x^+(0,t)$ for the second PDE subsystem, contrary to \cite{Hu2019siam}.
Thus, \eqref{eq:trafo1_kerneleq2_bc0} is obtained by using \eqref{eq:x0tilde} and requiring the coefficient of $x^+(0,t)$ to vanish:
\begin{multline}
	\label{eq:bilanz2}
	K^{+-}(z,0)\Lambda^-(0) \hspace{-.7cm} \underbrace{x^-(0,t)}_{Q^\text{R}x^+(0,t)+Q^\perp\tilde x_2^-(0,t)} \hspace{-.7cm} - K^{++}(z,0)\Lambda^+(0)x^+(0,t) \\
	= A_0^+(z)x^+(0,t) + B_0^+(z) \tilde x_2^-(0,t).
\end{multline}

\begin{lemma}[Volterra kernel]
	\label{lem:kernel}
	The kernel equations \eqref{eq:trafo1_kerneleq1}--\eqref{eq:trafo1_kerneleq2} admit a piecewise continuous solution $K(z,\zeta)$ on $\mathcal T$.
\end{lemma}

PROOF.
In \cite{Hu2019siam}, the existence of a piecewise continous solution $K^{--}(z,\zeta)$ and $K^{-+}(z,\zeta)$ of \eqref{eq:trafo1_kerneleq1} is shown.
This implies the same for \eqref{eq:trafo1_kerneleq2}, as these kernel equations can be traced back to \eqref{eq:trafo1_kerneleq1} by the simple substitutions $K^{+\pm}(z,\zeta) \to K^{-\mp}(z,\zeta)$, $\Lambda^\pm(z) \to -\Lambda^\mp(z)$, $A^{\pm-}(z) \to A^{\mp+}(z)$, $A^{\pm+}(z) \to A^{\mp-}(z)$, and $Q \to Q^\text{R}$.
\hfill $\square$

It is straight-forward to show that the transformation \eqref{eq:trafo1} is invertible because of its Volterra type.

In the end, \eqref{eq:trafo1} maps \eqref{eq:sys} into the form
\begin{subequations}
	\label{eq:sys_trafo1}
	\begin{align}
		\label{eq:sys_trafo1_PDEp}
		\partial_t \bar x^+(z,t) &= -\Lambda^+(z)\partial_z \bar x^+(z,t) + A_0^+(z) \bar x^+(0,t) \\
		&\hspace{3.2cm} + B_0^+(z) \tilde x_2^-(0,t) \nonumber\\
		\label{eq:sys_trafo1_RB0}
		\bar x^+(0,t) &= Q\bar x^-(0,t) \\
		\partial_t \bar x^-(z,t) &= \Lambda^-(z)\partial_z \bar x^-(z,t) + A_0^-(z) \bar x^-(0,t) \\
		\label{eq:sys_trafo1-bc1}
		\bar x^-(1,t) &= \bar u(t),
	\end{align}
\end{subequations}
where the elements of $B_0^+(z)$ (cf.\ the balance in \eqref{eq:bilanz2} w.r.t.\ $\tilde x_2^-(0,t)$) and those of the strictly lower triangular matrices $A_0^\pm(z)$ are defined by the solution $K(z,\zeta)$ of \eqref{eq:trafo1_kerneleq1} and \eqref{eq:trafo1_kerneleq2}:%
\begin{subequations}
	\label{eq:def_A0_B0}
	\begin{align}
		\left[A_0^-(z)\right]_{i>j} \!&=\!
		\left[-\! K^{-+}(z,0)\Lambda^+(0)Q \!+\! K^{--}(z,0)\Lambda^-(0)\right]_{i>j} \\
		\left[A_0^+(z)\right]_{i>j} \!&=\!
		\left[K^{+-}(z,0)\Lambda^-(0) Q^\text{R} \!-\! K^{++}(z,0)\Lambda^+(0)\right]_{i>j} \\
		B_0^+(z) &= K^{+-}(z,0)\Lambda^-(0)Q^\perp.
	\end{align}
\end{subequations}
For simplicity of the presentation, in \eqref{eq:sys_trafo1-bc1}, a new input $\bar u(t)$ is introduced via the state feedback
\begin{multline}
	u(t) = \tint_0^1 \left(K^{--}(1,\zeta)x^-(\zeta,t)+K^{-+}(1,\zeta)x^+(\zeta,t)\right)\,\dd\zeta \\
	- Rx^+(1,t) + \bar u(t).
\end{multline}
This does not affect the mapping into strict-feedback form.

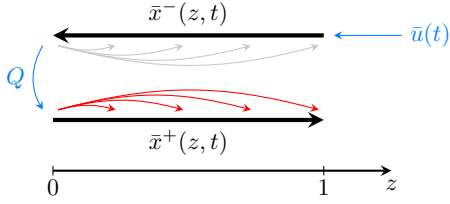
\begin{figure}
	\centering
	\input{figs/fig-system-trafo1.tex}
	\caption{Coupling structure of \eqref{eq:sys_trafo1} for $\Delta n=0$. Red arrows (\textcolor{red}{\rule[.5ex]{3ex}{.2ex}}) indicate a violation of the strict-feedback form due to $A_0^+(z)\bar x^+(0,t)$.}
	\label{fig:sys-trafo1}
\end{figure}

In \eqref{eq:sys_trafo1}, the equations are rearranged such that the transport equations with positive velocities, i.e., from $z=0$ to $z=1$, are written first.
This is done because the input in a strict-feedback form acts on the last state component, which in this case corresponds to $\bar x^-(z,t)$, specifically $\bar x^-(1,t)$.
However, because of $\bar x^+(0,t)$ in \eqref{eq:sys_trafo1_PDEp}, the dynamics \eqref{eq:sys_trafo1} is not in strict-feedback form, yet (cf.\ Fig.~\ref{fig:sys-trafo1}).
Therefore, next, a Fredholm-type integral transformation is used.

\begin{remark}
	\label{rem:a0}
	Uniqueness of the solution $K^{--}(z,\zeta)$ and $K^{-+}(z,\zeta)$ of \eqref{eq:trafo1_kerneleq1} requires additional BCs for the elements $[K^{--}(1,\zeta)]_{i>j}$ (see \cite{Hu2019siam}).
	Instead, it is also possible to assign the additional BCs
	\begin{equation}
		\left[K^{--}(z,0)\Lambda^-(0) - K^{-+}(z,0)\Lambda^+(0)Q\right]_{i>j} = 0
	\end{equation}
	for $z\in[\psi_i^-(\phi_i^-(1)-\phi_j^-(1)),1]$ such that $A_{0,ij}^-(z)=0$ holds for $z\in[\psi_i^-(\phi_i^-(1)-\phi_j^-(1)),1]$, with $\phi_i^-(z)=\tint_0^z \dd\zeta/\lambda_i^-(\zeta)$ and $\psi_i^-(\phi_i^-(z))=z$.
	The same applies in an analogous manner to the kernel equations \eqref{eq:trafo1_kerneleq2}.
\end{remark}

\subsection{Fredholm-type integral transformation}
\label{sec:fit}

The Fredholm-type integral transformation
\begin{equation}
	\label{eq:trafo2}
	\bar x^+(z,t) = \tilde x^+(z,t) + \tint_0^1 P_\text{I}(z,\zeta)\tilde x^+(\zeta,t)\,\dd\zeta
\end{equation}
is inherently invertible due to the kernel $P_\text{I}(z,\zeta)\in\Rset^{n_+\times n_+}$ being of strictly lower triangular form\footnote{A similar transformation of $\bar x^-(z,t)$ (instead of $\bar x^+(z,t)$ as in \eqref{eq:trafo2}) is used in \cite{Coron2017AUT} to achieve stabilization in minimum time.}.
If $P_\text{I}(z,\zeta)$ satisfies the kernel equations
\begin{subequations}
	\label{eq:trafo2_kerneleq}
	\begin{align}
		\Lambda^+(z) \partial_z P_\text{I}(z,\zeta) + \dd_\zeta(P_\text{I}(z,\zeta)\Lambda^+(\zeta)) &= 0 \\
		P_\text{I}(z,0)\Lambda^+(0) &= A_0^+(z) \\
		P_\text{I}(0,\zeta) &= 0,
	\end{align}
\end{subequations}
then \eqref{eq:trafo2} maps \eqref{eq:sys_trafo1} into the strict-feedback form
\begin{subequations}
	\label{eq:sys_trafo2}
	\begin{align}
		\label{eq:sys_trafo2-PDE1}
		\partial_t \tilde x^+(z,t) &= -\Lambda^+(z)\partial_z \tilde x^+(z,t) + \tilde A_0^+(z) \tilde x^+(1,t) \\
		&\hspace{3.2cm} + \tilde B_0^+(z) \tilde x_2^-(0,t) \nonumber \\
		\label{eq:sys_trafo2-BC1}
		\tilde x^+(0,t) &= Q\bar x^-(0,t) \\
		\label{eq:sys_trafo2-PDE2}
		\partial_t \bar x^-(z,t) &= \Lambda^-(z)\partial_z \bar x^-(z,t) + A_0^-(z) \bar x^-(0,t) \\
		\label{eq:sys_trafo2-BC2}
		\bar x^-(1,t) &= \bar u(t).
	\end{align}
\end{subequations}
Therein, $\tilde A_0^+(z) $ and $\tilde B_0^+(z)$ are defined by
\begin{subequations}
	\label{eq:def_A0_B0_tilde}
	\begin{align}
		\tilde A_0^+(z) + \tint_0^1 P_\text{I}(z,\zeta)\tilde A_0^+(\zeta)\,\dd\zeta &= P_\text{I}(z,1)\Lambda^+(1) \\
		\tilde B_0^+(z) + \tint_0^1 P_\text{I}(z,\zeta)\tilde B_0^+(\zeta)\,\dd\zeta &= B_0^+(z).
	\end{align}
\end{subequations}
The elements of the strictly lower triangular matrix $\tilde A_0^+(z)$ can be calculated successively, without the need to solve integral equations.
Using the method of characteristics, it is straight-forward to show that the kernel equations \eqref{eq:trafo2_kerneleq} admit a piecewise continuous solution $P_\text{I}(z,\zeta)$ on $[0,1]^2$.

Note that the transformation \eqref{eq:trafo2} and, thus, the mapping of the \eqref{eq:sys} into strict-feedback form is only made possible by the adjusted kernel equations \eqref{eq:trafo1_kerneleq2} that ensure a strictly lower triangular matrix $A_0^+(z)$ in \eqref{eq:sys_trafo1_PDEp}.



\subsection{Strict-feedback form}
\label{sec:sff}

\begin{figure}
	\centering
	\input{figs/fig-system-trafo2.tex}
	\caption{Strict-feedback form \eqref{eq:sys_trafo2} of the heterodirectional hyperbolic system \eqref{eq:sys}, highlighted by the arrows in blue (\textcolor{colorInput}{\rule[.5ex]{3ex}{.2ex}}), for the case $\Delta n=0$.}
	\label{fig:sys-trafo2}
\end{figure}
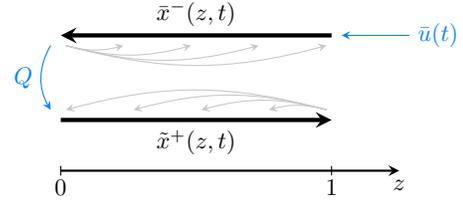

\begin{figure*}[t]
	\begin{subequations}
		\label{eq:sff}
		\begin{align}
			\label{eq:sff-x+}
			\partial_t x^+(z,t) &= -\Lambda^+(z)\partial_z  x^+(z,t) + F_1(z) x^+(1,t) + F_2(z) x^+(z,t) + \tint_z^1 F_3(z,\zeta) x^+(\zeta,t)\,\dd \zeta + G_1(z)(Q^\perp)^{\text{L}} x^-(0,t) \\
			x^+(0,t) &= Q x^-(0,t) + \tint_0^1 G_2(z)x^+(z,t)\,\dd z \\
			\label{eq:sff-x-}
			\partial_t x^-(z,t) &= \Lambda^-(z)\partial_z x^-(z,t) + F_4^-(z) x^-(0,t) + F_5(z) x^-(z,t) + \tint_0^z F_6(z,\zeta) x^-(\zeta,t)\,\dd\zeta \\
			&\hspace{6cm} + G_3(z) x^+(1,t) + G_4(z) x^+(z,t) + \tint_z^1 G_5(z,\zeta) x^+(\zeta,t)\,\dd\zeta  \nonumber \\
			\label{eq:sff-bc1}
			x^-(1,t) &= u(t)
		\end{align}
	\end{subequations}
\end{figure*}

To better understand the strict-feedback form of \eqref{eq:sys_trafo2}, first, assume $\Delta n=0$, which implies the absence of a matrix $\tilde B_0^+(z)$, and define an output $y(t)=\tilde x^+(1,t)$.
In view of this output, the first subsystem \eqref{eq:sys_trafo2-PDE1}--\eqref{eq:sys_trafo2-BC1} is in strict-feedback form as the PDEs propagate from $z=0$ to $z=1$.
The coupling is only in terms of the faster transport processes affecting the slower ones.
The (virtual) input to the first subsystem is the output $\bar x^-(0,t)$ of the second subsystem \eqref{eq:sys_trafo2-PDE2}--\eqref{eq:sys_trafo2-BC2}, where the transport direction is from $z=1$ to $z=0$.
It is again in strict-feedback form, with the control input $\bar u(t)$ at $z=1$.
By the properties of $Q$ ensured by Assumption~\ref{ass:Q} and the full boundary actuation via $u(t)$, the overall system is in strict-feedback form too, with a direct path $u(t) \to \bar x^-(z,t) \to \tilde x^+(z,t)$.
This is illustrated in Fig.~\ref{fig:sys-trafo2}.
In the asymmetric case, there are more control input components than there are transport equations with positive velocities, i.e., $n_->n_+\Leftrightarrow\Delta n>0$.
Still, $Q\bar x^-(0,t)=\tilde x_1^-(0,t)$ ensures full boundary actuation of the first subsystem, with additional control action via $\tilde x_2^-(0,t)$.

The following definition for a strict-feedback form of a heterodirectional hyperbolic system is suggested, which includes \eqref{eq:sys_trafo2} as a special case.

\begin{definition}[strict-feedback form]
	\label{def:sff}
	For a heterodirectional hyperbolic system with $\dim x^+(z,t)=n_+\le n_-=\dim x^-(z,t)$, input $u(t)\in\Rset^{n_-}$ and $\rank Q=n_+$, where $(Q^\perp)^{\text{L}}$ denotes the left inverse of a full-rank right annihilator $Q^\perp$ of $Q$, the representation \eqref{eq:sff} with lower triangular matrices $F_i$, $i=1,\dots,6$, and matrices $G_i$, $i=1,\dots,5$, without a special form is called a strict-feedback form.
\end{definition}

The strict-feedback form is particularly apparent when a spatial discretization is applied to the transport equations and the distributed state.
The input $u(t)\in\Rset^{n_-}$ can always be chosen such that \eqref{eq:sff-bc1} holds.

In \eqref{eq:sff}, $x^-(0,t)$ takes the role of an input w.r.t.\ the $n_+$ PDEs in \eqref{eq:sff-x+}.
The property $\rank Q=n_+$ ensures full actuation, with part of $x^-(0,t)$ directly actuating the boundary at $z=0$ and the remaining $\Delta n=n_--n_+\ge0$ components $(Q^\perp)^{\text{L}}x^-(0,t)$ acting distributed via $G_1(z)$.
In the original spirit of (ODE) backstepping, choosing a virtual stabilizing state feedback for the input $x^-(0,t)$ that drives $x^+(z,t)$ to zero implies a state transformation of the form $\bar x^-(z,t)=x^-(z,t) + \tint_0^1 P(z,\zeta)x^+(\zeta,t)\,\dd\zeta$.
Then, the second and final step is mainly concerned with setting $u(t)$, the control action and input to the $n_-$ PDEs in \eqref{eq:sff-x-}, such that the error $\bar x^-(z,t)$ goes to zero, which in turn ensures that the virtual feedback for $x^-(0,t)$ holds.
Both in the case of the virtual feedback for $x^-(0,t)$ and the choice of the controller $u(t)$, the design is facilitated by a prior Volterra integral transformation of $x^+(z,t)$ and $\bar x^-(z,t)$, respectively, as well as, potentially, the use of Lyapunov functionals.

Note that the same design procedure applies to systems in strict-feedback form with more than two subsystems.
Then, the input to the subsystem \eqref{eq:sff-x-}--\eqref{eq:sff-bc1} is the output of another transport process from $z=0$ to $z=1$.

\begin{remark}
	From a flatness-based perspective, $\tilde x^+(1,t)\in\Rset^{n_+}$ and $\tilde x_2^-(0,t)\in\Rset^{\Delta n}$ constitute a flat output $y(t)\in\Rset^{n_-}$ of the system \eqref{eq:sff}.
	This allows for a parametrization of all system variables $\tilde x^+(z,t)$, $\bar x^-(z,t)$ and $\bar u(t)$ in terms of $y(t)$ and predictions thereof, and a very simple control design.
\end{remark}

\section{Strict-feedback form for PDE-ODE systems}
\label{sec:po}

There are numerous backstepping designs for the stabilization of bidirectionally coupled hyperbolic PDE-ODE systems, where a finite-dimensional dynamics is present at the unactuated boundary of the PDE \eqref{eq:sys} (e.g., \cite{Zhou2012,DiMeglio2018aut,Deutscher2019IJC,Gehring2021MTNS}).
However, none of the approaches draw on a strict-feedback form, which is the basis for backstepping in the finite-dimensional case.
Although \cite{Gehring2021MTNS} suggests to use just such a form for a recursive control design, it is clear from the structure of the target system that this cannot be the case.
Therefore, this section addresses the strict-feedback form for PDE-ODE systems.
For simplicity, the two PDE subsystems are assumed to consist of pure transport equations only.

\subsection{Problem statement}

The interconnection of the heterodirectional hyperbolic PDE \eqref{eq:sys} without in-domain coupling, i.e., $A^{\pm\pm}(z)=0$, and an ODE at the unactuated boundary at $z=0$ results in the bidirectionally coupled PDE-ODE system
\begin{subequations}
	\label{eq:po:sys}
	\begin{align}
		\label{eq:po:sys-ode}
		\dot\xi(t) &= F\xi(t) + B x^-(0,t) \\
		\label{eq:po:sys-pde}
		\begin{bmatrix} \partial_t x^-(z,t) \\ \partial_t x^+(z,t) \end{bmatrix} &= \begin{bmatrix} \Lambda^-(z) & 0 \\ 0 & -\Lambda^+(z) \end{bmatrix} \begin{bmatrix} \partial_z x^-(z,t) \\ \partial_z x^+(z,t) \end{bmatrix} \\
		\label{eq:po:sys-bc0}
		x^+(0,t) &= Qx^-(0,t) + C\xi(t) \\
		\label{eq:po:sys-bc1}
		x^-(1,t) &= Rx^+(1,t) + u(t).
	\end{align}
\end{subequations}
The state $\xi(t)\in\Rset^{n_0}$ of the ODE subsystem \eqref{eq:po:sys-ode} affects the PDE subsystem \eqref{eq:po:sys-pde}--\eqref{eq:po:sys-bc1} via $C\in\Rset^{n_+\times n_0}$, while the boundary value $x^-(0,t)$ via $B\in\Rset^{n_0\times n_-}$ takes the role of an input w.r.t.\ \eqref{eq:po:sys-ode}.
In addition to the exact controllability of the PDE subsystem (recall Assumption~\ref{ass:Q}), it is assumed that the ODE subsystem \eqref{eq:po:sys-ode} is controllable.

\begin{assumption}
	\label{ass:FBpair}
	The matrix pair $(F,B)$ is controllable.
\end{assumption}

Both conditions are necessary for the existence of a strict-feedback form.
It is apparent from the structural representation in Fig.~\ref{fig:posys} that the PDE-ODE system \eqref{eq:po:sys} with its three subsystems is not in strict-feedback form.
It would posses the special form if $x^+(1,t)$ instead of $x^-(0,t)$ in \eqref{eq:po:sys-ode}.
Similar to Section~\ref{sec:fit}, next, a transformation is used to map \eqref{eq:po:sys} into strict-feedback form.

\begin{figure}
	\centering
	\input{figs/fig-POsystem.tex}
	\caption{Coupling structure of the PDE-ODE system \eqref{eq:po:sys}. The arrows in blue (\textcolor{colorInput}{\rule[.5ex]{3ex}{.2ex}}) highlight actions related to controllability.}
	\label{fig:posys}
\end{figure}

\subsection{Transformation into strict-feedback form}

Consider the invertible transformation
\begin{equation}
	\label{eq:po:trafo}
	\bar\xi(t) = \xi(t) - \tint_0^1 N(z) x^+(z)\,\dd z
\end{equation}
with kernel $N(z)\in\Rset^{n_0\times n_+}$, $z\in[0,1]$.
If $N(z)$ satisfies the matrix-valued initial value problem
\begin{subequations}
	\label{eq:po:trafo-ivp}
	\begin{align}
		\dd_z(N\Lambda^+)(z) &= (F-BQ^\text{R}C)N(z) \\
		(N\Lambda^+)(0) &= BQ^\text{R},
	\end{align}
\end{subequations}
then \eqref{eq:po:trafo} maps \eqref{eq:po:sys} into the strict-feedback form
\begin{subequations}
	\label{eq:po:sys-trafo}
	\begin{align}
		\label{eq:po:sys-trafo-ode}
		\dot{\bar\xi}(t) &= \bar F\bar\xi(t) + \bar Bx^+(1,t) + BQ^\perp \tilde x_2^-(0,t) \\
		\partial_t x^+(z,t) &= -\Lambda^+(z)\partial_z x^+(z,t) \\
		x^+(0,t) &= Qx^-(0,t) + C\bar\xi(t) + \tint_0^1 CN(z)x^+(z,t)\,\dd z \\
		\partial_t x^-(z,t) &= \Lambda^-(z)\partial_z x^-(z,t) \\
		x^-(1,t) &= Rx^+(1,t) + u(t)
	\end{align}
\end{subequations}
with $\bar F=F-BQ^\text{R}C$ and $\bar B=(N\Lambda^+)(1)$.
The solution of \eqref{eq:po:trafo-ivp} is straight-forward.
It can be determined by considering an eigenvalue problem for $\bar F$ and solving $n_0$ vector-valued initial value problems.
Based on that, the Hautus test allows to verify that the controllability of the ODE \eqref{eq:po:sys-trafo-ode} is preserved under the transformation \eqref{eq:po:trafo}.

\begin{lemma}[controllability of ODE]
	\label{lem:controllabilityODE}
	If Assumption~\ref{ass:FBpair} holds, then the ODE \eqref{eq:po:sys-trafo-ode} with the matrix pair $(\bar F,[\bar B,BQ^\perp])$ is controllable.
\end{lemma}

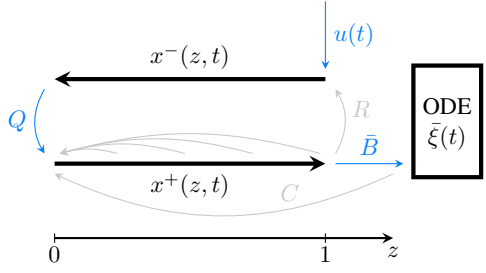
\begin{figure}
	\centering
	\input{figs/fig-POsystem-transformed.tex}
	\caption{Strict-feedback form \eqref{eq:po:sys-trafo} of the PDE-ODE system \eqref{eq:po:sys}, highlighted by the arrows in blue (\textcolor{colorInput}{\rule[.5ex]{3ex}{.2ex}}), for the case $\Delta n=0$.}
	\label{fig:posys-trafo}
\end{figure}

The transformation \eqref{eq:po:trafo} is strongly related to an Artstein transformation for linear time-delay systems (see, e.g., \cite{Artstein1982tac}).
In fact, \eqref{eq:po:sys-trafo} essentially moves the ODE through the PDE subsystem with $x^+(z,t)$, meaning that $\bar\xi(t)$ corresponds to a delayed state of $\xi(t)$.
This is apparent when comparing the structure of \eqref{eq:po:sys} in Fig.\ \ref{fig:posys} to that of \eqref{eq:po:sys-trafo} in Fig.\ \ref{fig:posys-trafo}.
In the latter, the system is in strict-feedback form as there is a direct path $u(t)\to x^-(z,t)\to x^+(z,t)\to\bar\xi(t)$ of actuation, with all remaining couplings in the opposite direction only.
Note that this is only possible due to Assumptions~\ref{ass:Q} and \ref{ass:FBpair}, and the full boundary actuation via the control input $u(t)$.

\section{Concluding remarks}

The strict-feedback form introduced for heterodirectional hyperbolic PDEs and PDE-ODE systems is only a first step in better understanding and, consequently, extending backstepping results for hyperbolic distributed-parameter systems.
The form directly suggests a large class of coupled systems for which a stabilizing controller can be designed recursively, i.e, in the original spirit of backstepping, by potentially using Lyapunov functionals.
Importantly, both the design and the system structure are linked to controllability properties.
This also allows to draw parallels to the well-known concepts of input-state and input-output linearization as well as zero dynamics for finite-dimensional systems.


\color{black}

\section*{Acknowledgment}
This research was funded in part by the Austrian Science Fund (FWF) [I 6519-N]. The author also thanks Dr.\ Abdurrahman Irscheid of Saarland University, Germany, for many valuable discussions and a very insightful suggestion.





\bibliographystyle{plain}
\bibliography{/home/ngehring/Documents/90_Literatur/mybib}

\end{document}

%% file: figs/fig-system.tex
\begin{tikzpicture}[scale=.9, transform shape]
	\normalsize
	\pgfmathsetmacro{\zAxis}{5}
	\pgfmathsetmacro{\zLeftBoundary}{4/5*\zAxis}
	\pgfmathsetmacro{\Tick}{0.07}
	\pgfmathsetmacro{\UpperChannel}{2}
	\pgfmathsetmacro{\LowerChannel}{0.75}
	\pgfmathsetmacro{\inDomain}{1/6*\zAxis}
	\pgfmathsetmacro{\out}{0.15}
	\pgfmathsetmacro{\biODE}{1/5*\zAxis}
	\pgfmathsetmacro{\ODEwidth}{1/5*\zAxis}
	\pgfmathsetmacro{\ODEextraheight}{0.2}
    \pgfmathsetmacro{\myFactor}{1.5}
	
	\draw [->,thick] (0,0) coordinate (B0) node [below] {$0$}
	|- (\zLeftBoundary,0) coordinate (B1) node [below] {$1$} 
	|- (\zAxis,0) node (xaxis) [below] {$z$};
	
	\draw ($ (B0)+(0,\Tick) $) -- ($ (B0)-(0,\Tick) $);
	\draw ($ (B1)+(0,\Tick) $) -- ($ (B1)-(0,\Tick) $);
	
	\draw [ultra thick,->]
	($ (B1)+(0,\UpperChannel) $) coordinate (B1_x1)
	--
	($ (B0)+(0,\UpperChannel) $) coordinate (B0_x1)
	node[pos=0.5,above]{$x^-(z,t)$};
	
	\draw [ultra thick,->]
	($ (B0)+(0,\LowerChannel) $) coordinate (B0_x2)
	--
	($ (B1)+(0,\LowerChannel) $) coordinate (B1_x2)
	node[pos=0.5,below]{$x^+(z,t)$}
	;
	
	\foreach \i in {1,...,4}
	{
		\draw [<->] ($ (B0_x2)+(\i*\inDomain,\out) $) -- ($ (B0_x1)+(\i*\inDomain,-\out) $);
	}
	
	\draw [->]
	($ (B1_x2)+(\out,\out) $) to[bend right] node[right]{$R$} ($ (B1_x1)+(\out,-\out) $);
	\draw [->,colorInput]
	($ (B0_x1)-(\out,\out) $) to[bend right] node[left]{$Q$} ($ (B0_x2)-(\out,-\out) $);
	
	\draw [<-,colorInput] ($ (B1_x1)+(\out,0) $) -- ($ (B1_x1)+(\out+\biODE,0) $)
	node[right]{$u(t)$};
	
\end{tikzpicture}

%% file: figs/fig-system-trafo1.tex
\begin{tikzpicture}[scale=.9, transform shape]
	\normalsize
	\pgfmathsetmacro{\zAxis}{5}
	\pgfmathsetmacro{\zLeftBoundary}{4/5*\zAxis}
	\pgfmathsetmacro{\Tick}{0.07}
	\pgfmathsetmacro{\UpperChannel}{2}
	\pgfmathsetmacro{\LowerChannel}{0.75}
	\pgfmathsetmacro{\inDomain}{1/6*\zAxis}
	\pgfmathsetmacro{\out}{0.15}
	\pgfmathsetmacro{\biODE}{1/5*\zAxis}
	\pgfmathsetmacro{\ODEwidth}{1/5*\zAxis}
	\pgfmathsetmacro{\ODEextraheight}{0.2}
    \pgfmathsetmacro{\myFactor}{1.5}
	
	\draw [->,thick] (0,0) coordinate (B0) node [below] {$0$}
	|- (\zLeftBoundary,0) coordinate (B1) node [below] {$1$} 
	|- (\zAxis,0) node (xaxis) [below] {$z$};
	
	\draw ($ (B0)+(0,\Tick) $) -- ($ (B0)-(0,\Tick) $);
	\draw ($ (B1)+(0,\Tick) $) -- ($ (B1)-(0,\Tick) $);
	
	\draw [ultra thick,->]
	($ (B1)+(0,\UpperChannel) $) coordinate (B1_x1)
	--
	($ (B0)+(0,\UpperChannel) $) coordinate (B0_x1)
	node[pos=0.5,above]{$\bar x^-(z,t)$};
	
	\draw [ultra thick,->]
	($ (B0)+(0,\LowerChannel) $) coordinate (B0_x2)
	--
	($ (B1)+(0,\LowerChannel) $) coordinate (B1_x2)
	node[pos=0.5,below]{$\bar x^+(z,t)$}
	;
	
	\draw [->,colorInput]
	($ (B0_x1)-(\out,\out) $) to[bend right] node[left]{$Q$} ($ (B0_x2)-(\out,-\out) $);
	
	\draw [<-,colorInput] ($ (B1_x1)+(\out,0) $) -- ($ (B1_x1)+(\out+\biODE,0) $)
	node[right]{$\bar u(t)$};
	
	\foreach \xi in {1,2,3,4}
		\draw[->,red] ($ (B0_x2)+(\out/2,\out) $) to[bend left=15] ($ (B0_x2)+(\zLeftBoundary*\xi/4-\out/2,\out) $);
		
	\foreach \xi in {1,2,3,4}
		\draw[->,colorODE] ($ (B0_x1)+(\out/2,-\out) $) to[bend right=15] ($ (B0_x1)+(\zLeftBoundary*\xi/4-\out/2,-\out) $);
	
\end{tikzpicture}

%% file: figs/fig-system-trafo2.tex
\begin{tikzpicture}[scale=.9, transform shape]
	\normalsize
	\pgfmathsetmacro{\zAxis}{5}
	\pgfmathsetmacro{\zLeftBoundary}{4/5*\zAxis}
	\pgfmathsetmacro{\Tick}{0.07}
	\pgfmathsetmacro{\UpperChannel}{2}
	\pgfmathsetmacro{\LowerChannel}{0.75}
	\pgfmathsetmacro{\inDomain}{1/6*\zAxis}
	\pgfmathsetmacro{\out}{0.15}
	\pgfmathsetmacro{\biODE}{1/5*\zAxis}
	\pgfmathsetmacro{\ODEwidth}{1/5*\zAxis}
	\pgfmathsetmacro{\ODEextraheight}{0.2}
    \pgfmathsetmacro{\myFactor}{1.5}
	
	\draw [->,thick] (0,0) coordinate (B0) node [below] {$0$}
	|- (\zLeftBoundary,0) coordinate (B1) node [below] {$1$} 
	|- (\zAxis,0) node (xaxis) [below] {$z$};
	
	\draw ($ (B0)+(0,\Tick) $) -- ($ (B0)-(0,\Tick) $);
	\draw ($ (B1)+(0,\Tick) $) -- ($ (B1)-(0,\Tick) $);
	
	\draw [ultra thick,->]
	($ (B1)+(0,\UpperChannel) $) coordinate (B1_x1)
	--
	($ (B0)+(0,\UpperChannel) $) coordinate (B0_x1)
	node[pos=0.5,above]{$\bar x^-(z,t)$};
	
	\draw [ultra thick,->]
	($ (B0)+(0,\LowerChannel) $) coordinate (B0_x2)
	--
	($ (B1)+(0,\LowerChannel) $) coordinate (B1_x2)
	node[pos=0.5,below]{$\tilde x^+(z,t)$}
	;
	
	\draw [->,colorInput]
	($ (B0_x1)-(\out,\out) $) to[bend right] node[left]{$Q$} ($ (B0_x2)-(\out,-\out) $);
	
	\draw [<-,colorInput] ($ (B1_x1)+(\out,0) $) -- ($ (B1_x1)+(\out+\biODE,0) $)
	node[right]{$\bar u(t)$};
	
	\foreach \xi in {1,2,3,4}
		\draw[->,colorODE] ($ (B0_x1)+(\out/2,-\out) $) to[bend right=15] ($ (B0_x1)+(\zLeftBoundary*\xi/4-\out/2,-\out) $);
	\foreach \xi in {1,2,3,4}
		\draw[->,colorODE] ($ (B1_x2)+(-\out/2,\out) $) to[bend right=15] ($ (B1_x2)-(\zLeftBoundary*\xi/4-\out/2,-\out) $);
	
\end{tikzpicture}

%% file: figs/fig-POsystem.tex
\begin{tikzpicture}[scale=.9, transform shape]
	\normalsize
	\pgfmathsetmacro{\zAxis}{5}
	\pgfmathsetmacro{\zLeftBoundary}{4/5*\zAxis}
	\pgfmathsetmacro{\Tick}{0.07}
	\pgfmathsetmacro{\UpperChannel}{2}
	\pgfmathsetmacro{\LowerChannel}{0.75}
	\pgfmathsetmacro{\inDomain}{1/6*\zAxis}
	\pgfmathsetmacro{\out}{0.15}
	\pgfmathsetmacro{\biODE}{1/5*\zAxis}
	\pgfmathsetmacro{\ODEwidth}{1/5*\zAxis}
	\pgfmathsetmacro{\ODEextraheight}{0.2}
    \pgfmathsetmacro{\myFactor}{1.5}
	
	\draw [->,thick] (0,0) coordinate (B0) node [below] {$0$}
	|- (\zLeftBoundary,0) coordinate (B1) node [below] {$1$} 
	|- (\zAxis,0) node (xaxis) [below] {$z$};
	
	\draw ($ (B0)+(0,\Tick) $) -- ($ (B0)-(0,\Tick) $);
	\draw ($ (B1)+(0,\Tick) $) -- ($ (B1)-(0,\Tick) $);
	
	\draw [ultra thick,->]
	($ (B1)+(0,\UpperChannel) $) coordinate (B1_x1)
	--
	($ (B0)+(0,\UpperChannel) $) coordinate (B0_x1)
	node[pos=0.5,above]{$x^-(z,t)$};
	
	\draw [ultra thick,->]
	($ (B0)+(0,\LowerChannel) $) coordinate (B0_x2)
	--
	($ (B1)+(0,\LowerChannel) $) coordinate (B1_x2)
	node[pos=0.5,below]{$x^+(z,t)$}
	;
	
	\draw [->]
	($ (B1_x2)+(\out,\out) $) to[bend right] node[right]{$R$} ($ (B1_x1)+(\out,-\out) $);
	\draw [->,colorInput]
	($ (B0_x1)-(\out,\out) $) to[bend right] node[left, pos=.3]{$Q$} ($ (B0_x2)-(\out,-\out) $);
	
	\draw [<-,colorInput] ($ (B1_x1)+(\out,0) $) -- ($ (B1_x1)+(\out+\biODE,0) $)
	node[right]{$u(t)$};
	
	\draw [->,colorInput] ($ (B0_x1)-(\out,0) $) -- ($ (B0_x1)-(\out+\biODE,0) $)
	coordinate (ODEup) node[pos=0.5,above]{$B$};
	\draw [->] ($ (B0_x2)-(\out+\biODE,0) $) coordinate (ODEdown)
	-- ($ (B0_x2)-(\out,0) $) node[pos=0.5,above]{$C$};
	\draw[ultra thick] ($ (ODEup)-(\out,-\ODEextraheight) $) rectangle coordinate (ODE) ++(-\ODEwidth,-2*\ODEextraheight-\UpperChannel+\LowerChannel) node[midway](ODE){};
	\draw
	node at ($(ODE) + (0,\myFactor*\out)$) {ODE}
    node at ($(ODE) + (\ODEwidth/2,0)$) (ODEwest) {}
	node at ($(ODE) - (0,\myFactor*\out)$) {$\xi(t)$}
	;
	
\end{tikzpicture}

%% file: figs/fig-POsystem-transformed.tex
\begin{tikzpicture}[scale=.9, transform shape]
	\normalsize
	\pgfmathsetmacro{\zAxis}{5}
	\pgfmathsetmacro{\zLeftBoundary}{4/5*\zAxis}
	\pgfmathsetmacro{\Tick}{0.07}
	\pgfmathsetmacro{\UpperChannel}{2.35}
	\pgfmathsetmacro{\LowerChannel}{1.1}
	\pgfmathsetmacro{\inDomain}{1/6*\zAxis}
	\pgfmathsetmacro{\out}{0.15}
	\pgfmathsetmacro{\biODE}{1/5*\zAxis}
	\pgfmathsetmacro{\ODEwidth}{1/5*\zAxis}
	\pgfmathsetmacro{\ODEextraheight}{0.2}
    \pgfmathsetmacro{\myFactor}{1.5}
	
	\draw [->,thick] (0,0) coordinate (B0) node [below] {$0$}
	|- (\zLeftBoundary,0) coordinate (B1) node [below] {$1$} 
	|- (\zAxis,0) node (xaxis) [below] {$z$};
	
	\draw ($ (B0)+(0,\Tick) $) -- ($ (B0)-(0,\Tick) $);
	\draw ($ (B1)+(0,\Tick) $) -- ($ (B1)-(0,\Tick) $);
	
	\draw [ultra thick,->]
	($ (B1)+(0,\UpperChannel) $) coordinate (B1_x1)
	--
	($ (B0)+(0,\UpperChannel) $) coordinate (B0_x1)
	node[pos=0.5,above]{$x^-(z,t)$};
	
	\draw [ultra thick,->]
	($ (B0)+(0,\LowerChannel) $) coordinate (B0_x2)
	--
	($ (B1)+(0,\LowerChannel) $) coordinate (B1_x2)
	node[pos=0.5,below]{$x^+(z,t)$}
	;
	
	\draw [->,colorODE]
	($ (B1_x2)+(\out,\out) $) to[bend right] node[right, pos=.7]{$R$} ($ (B1_x1)+(\out,-\out) $);
	\draw [->,colorInput]
	($ (B0_x1)-(\out,\out) $) to[bend right] node[left]{$Q$} ($ (B0_x2)-(\out,-\out) $);
	
	\draw [<-,colorInput] ($ (B1_x1)+(0,\out) $) -- node[right]{$u(t)$} ($ (B1_x1)+(0,\out+\biODE) $);
	
	\draw [<-,white] ($ (B1_x1)+(\out,0) $) -- ($ (B1_x1)+(\out+\biODE,0) $)
	coordinate (ODEup);
	\draw [<-,colorInput] ($ (B1_x2)+(\out+\biODE,0) $) coordinate (ODEdown)
	-- ($ (B1_x2)+(\out,0) $) node[pos=0.5,above]{$\bar B$};
	\draw[ultra thick] ($ (ODEup)+(\out,\ODEextraheight) $) rectangle coordinate (ODE) ++(\ODEwidth,-2*\ODEextraheight-\UpperChannel+\LowerChannel) node[midway](ODE){};
	\draw[->,colorODE]
	($ (ODEdown)-(\out,\out) $) to[bend left=25] node[above, pos=.3]{$C$} ($ (B0_x2)+(-0*\out,-\out) $);
	\foreach \xi in {1,2,3,4}
		\draw[->,colorODE] ($ (B0_x2)+(\zLeftBoundary*\xi/4-\out/2,\out) $) to[bend right=15] ($ (B0_x2)+(\out/2,\out) $);
	\draw
	node at ($(ODE) + (0,\myFactor*\out)$) {ODE}
    node at ($(ODE) + (\ODEwidth/2,0)$) (ODEwest) {}
	node at ($(ODE) - (0,\myFactor*\out)$) {$\bar\xi(t)$};
	
\end{tikzpicture}